\documentclass{elsart}

\usepackage{graphics}

\usepackage{amssymb}
\def\A{{\mathcal A}}
\def\H{{\mathcal H}}
\def\K{{\mathcal K}}
\def\D{{\mathcal D}}
\def\B{{\mathcal B}}
\def\trip{{|\!|\!|}}
\def\End{\mbox{\rm End}}

\def\vect{\overrightarrow}
\def\bra{\langle}
\def\ket{\rangle}
\def\tr{\mbox{\rm Tr}}
\def\SS{{\mathcal S}}
\def\bea{\begin{eqnarray}}
\def\eea{\end{eqnarray}}

\def\be{\begin{equation}}
\def\ee{\end{equation}}
\def\og{{``}}
\def\fg{{''}}

\begin{document}

\begin{frontmatter}


\title{Canonical quantization and the spectral action, a nice example}
\author{Fabien Besnard}
\ead{Fabien.Besnard@wanadoo.fr}
\address{EPF, 3 bis rue Lakanal, 92330 Sceaux, FRANCE}

\title{}


\author{}

\address{}

\begin{abstract}
We study the canonical quantization of the theory given by Chamseddine-Connes spectral action on a particular finite spectral triple with algebra $M_2(\Cset)\oplus\Cset$. We define a quantization of the natural distance associated with this noncommutative space and show that the quantum distance operator has a discrete spectrum. We also show that it would be the same for any other geometric quantity. Finally we propose a physical Hilbert space for the quantum theory. This spectral triple had been previously considered by Rovelli as a toy model, but with a different action which was not gauge-invariant. The results are similar in both cases, but the gauge-invariance of the spectral action manifests itself by the presence of a non-trivial degeneracy structure for our distance operator. 
 \end{abstract}

\end{frontmatter}

\section{Introduction}
One of the great successes of noncommutative geometry is the computation of the standard model action coupled to euclidean gravity from an almost commutative spectral triple on which is defined the Chamseddine-Connes action \cite{cc} :

\be
S(\D)=\bra \psi|\D\psi\ket+\tr(\chi({\D^2\over m_0^2}))\label{ccaction}
\ee

where $\D$ is the Dirac operator, $\psi$ the spinor field, $\chi$ a smooth approximation of the characteristic function of $[0;1]$ and $m_0$ is a mass scale. The very appealing feature of this action is that it treats gravity and other forces on the same footing, the gauge bosons of the standard model together with the Higgs field appearing naturally as inner perturbations of the Dirac operator $\D\rightarrow \D+A+JAJ^{-1}$. The action (\ref{ccaction}) only depends on the eigenvalues of the Dirac operator, as required by the spectral action principle put forward in \cite{cc}. This principle is a generalization of diffeomorphism invariance which can be naturally formulated in the algebraic language of noncommutative geometry.

Several quantization procedures are possible. The renormalization group approach has been most intensively studied, in particular by Connes. It has recently led to a prediction for the Higgs' mass (under the \og big desert\fg\  hypothesis, see \cite{connesmarcolli}). However non-perturbative attempts also exist, namely canonical quantization \cite{rovelli}, and path-integral quantization \cite{hale}. See also the intriguing approach of \cite{strup} where the authors use mathematical tools from Loop Quantum Gravity. 

In this paper we will focus on the canonical quantization of a toy model directly taken from \cite{rovelli}. In the latter paper the author succeeded in canonically quantizing the theory defined by the action $\tr(DMD)$ where $M$ is a matrix with null determinant whose purpose is to get a non-trivial space of solutions, since the extremum of the quadratic action $S(D)=\tr(D^2)$ is trivially $D=0$. However this does not follow the spectral action principle\footnote{However, in \cite{hale} it is shown that extremizing the action of \cite{rovelli} is equivalent to extremizing the spectral invariant action $\tr(\tilde D^2)$ on a submanifold of the configuration space, where $\tilde D$ in an effective Dirac operator constructed out of $D$.}, and as the author points out in \cite{rovelli} it is crucial to extend the explorations in that direction. Path quantization on several finite spectral triples for the action $\tr(D^2)$  have been worked out in \cite{hale}. Canonical quantization would not work in this setting for the already quoted fact that the phase space is trivial. Instead we will use the spectral action of Chamsedinne-Connes (without matter) and look for the canonical quantization of the theory it defines on the finite spectral triple considered in \cite{rovelli}. In particular we will show that distance is a well-defined observable in the quantum theory, with a discrete spectrum. This is in complete agreement with the results of \cite{rovelli} and \cite{hale}, although they are obtained in a slightly different context.

The paper is organized as follows : in section 2 we recall the definitions of the finite spectral triple and of the spectral action we use. In section 3 we compute the spectrum of the square of the Dirac operator, and use it to compute the spectral action. We also determine the phase space. In section 4 we use canonical quantization to define a Hilbert space for the quantum theory which bears a unitary representation of the gauge group. In section 5 we recall the precise definition and formula for the classical distance and construct its quantized version. In section 6 we propose a physical Hilbert space in which states are gauge-invariant. Section 7 contains discussion and outlook. Finally, in the appendix we prove some mathematical results we need about our finite spectral triple.

We use Planck units throughout the paper.

\section{Definitions}
We recall here for convenience the definition of the spectral triple $(\A,\H,\D)$ studied in \cite{rovelli}.

\begin{defn}
The $C^*$-algebra is $\A=M_2(\Cset)\oplus\Cset$, the Hilbert space is $\H=M_3(\Cset)$, with the scalar product defined by : $\forall \psi,\psi'\in \H$, $\bra \psi|\psi'\ket=\tr(\psi^*\psi')$. 

The representation of $\A$ in $\H$ is given by $\pi : \A\longrightarrow \End(\H)$, $\pi(a)\psi=\pi(M,\alpha)\psi=\pmatrix{M&0\cr 0&\alpha}\psi$ with $M\in M_2(\Cset)$, $\alpha\in\Cset$.
\end{defn}

The real structure is $J(\psi)=\psi^*$. There are also another operator $\chi$ on$\H$ which is called chirality and plays the role of the volume form. Since we won't make use of it in the core of this paper we refer to the appendix for its definition.

For any $m=\pmatrix{m_1\cr m_2}$, $m_1,m_2\in\Cset$, let us write $A(m)=\pmatrix{0_2& m\cr m^*&0}$ where $0_2$ is the null $2\times 2$ matrix. As proven in \cite {kraj}, for any such $m$ there is a Dirac operator on $\H$ defined by $\D_m\psi=A(m)\psi+\psi A(m)^*$, that is to say :
\be
\D_m=\Delta(m)+J\Delta(m)J^{-1}\label{dm}
\ee
where $\Delta(m) : \H\longrightarrow \H$ is the left multiplication by $A(m)$.

It is this class of Dirac operators that is used in \cite{rovelli} and that we will mainly consider in this paper. We think a clarification is needed here : this is not the most general Dirac operator on $(\A,\H)$ (see appendix A, where we also show that the sector of the theory that we consider here can be viewed as a gauge theory compatible with a given fixed structure on the base space). However, to allow comparison with previous results we will stick to this simple form for Dirac operators. It must be said that another reason is that we are not yet able to deal with the most general case, even at the classical level (see appendix B). 

For the sake of simplicity we will often write $A$ instead of $A(m)$. Note that our $A$ plays the role of the D in \cite{rovelli} which should not be confused with $\D$.

We now define the spectral action we will use throughout the paper.

\begin{defn}
The spectral action is (in Planck units) :
$$S(m)=\tr(\chi\Big({\D_m^2\over m_0^2}\Big))$$
where $m_0^2\in]0;+\infty[$ is a cut-off parameter, and $\chi$ a suitable approximation of the characteristic function of $[0;1]$.
This action is defined on the configuration space ${\mathcal C}$ of all Dirac operators on $\H$ of the form (\ref{dm}), which is identified to $\Cset^2$ with coordinate $m$. 
\end{defn}

Roughly, the spectral action just counts the number of eigenvalues of $\D_m^2$ which are below the cut-off.

\section{Computation of the spectral action}

Let $\lambda$ be an eigenvalue of $\D_m^2$ and $\psi$ an associated eigenvector. Thus :

\be
\D_m^2\psi=A^2\psi+2A\psi A+\psi A^2=\lambda\psi\label{eigen}
\ee

Since $A=A^*$ it is diagonalizable. Write $A=PBP^{-1}$ with $B$ a diagonal matrix, and write $\phi=P^{-1}\psi P$. Then (\ref{eigen}) is equivalent to :

\be
B^2\phi+2B\phi B+\phi B^2=\lambda \phi\label{diago}
\ee

with $B=$diag$(\lambda_1,\lambda_2,\lambda_3)$, (\ref{diago}) gives for all $i,j$ in $\{1;2;3\}$ :

\be
[(\lambda_i+\lambda_j)^2-\lambda]\phi_{ij}=0
\ee

Thus $Sp(\D_m^2)=\{(\lambda_i+\lambda_j)^2|\lambda_i,\lambda_j\in Sp(A)\}$. But we must look for the multiplicities. It is clear that $4\lambda_i^2$ is of multiplicity 1, whereas $(\lambda_i+\lambda_j)^2$ is of multiplicity 2 for $i\not=j$, but we must add the multiplicities if it happens that some values of $(\lambda_i+\lambda_j)^2$ coincide. In fact it is easy to calculate the $\lambda_i$ in term of $m$. One finds :

$$\lambda_1=0,\quad \lambda_2=-\|m\|,\quad \lambda_3=\|m\|$$

So the eigenvalues of $\D_m^2$ are : $0$ with multiplicity $3=1+2$, $\|m\|^2$ with multiplicity $4=2+2$, and $4\|m\|^2$ with multiplicity $2=1+1$.

As a direct application of the above, one gets :

\be
S(m)=3+4\chi({\|m\|^2\over m_0^2})+2\chi({4\|m\|^2\over m_0^2})
\ee

Let us suppose that $\chi(x)=1$ for $x\leq 1-\delta$ and $\chi(x)=0$ for $x\geq 1$. Then one has $S(m)=3$ for $\|m\|\geq m_0$, $S(m)=7$ for ${m_0\over 2}\leq \|m\|<(1-\delta)^{1/2}m_0$ and $S(m)=9$ for $\|m\|<(1-\delta)^{1/2}{m_0\over 2}$, with a smooth interpolation between each domain where $S$ is constant.


The equations of motion are obtained by extremizing the action with respect to $m$. It is clear that every Dirac operator is a solution to these equations except for those whose parameter $m$ falls in the small shells where $S$ is not constant. The phase space $\Gamma$, which is the space of solutions of the equations of motions\footnote{We use the terminology of \cite{rovelli} here. Sometimes the phase space is called the space of motion when one adopts this point of view.}, has a structure which depends on $\delta$, but the latter can be chosen as small as one wishes and has no physical importance. Thus $\Gamma$ can be taken to be :
$$\Gamma=\Gamma_1\sqcup\Gamma_2\sqcup\Gamma_3$$
It has three components : $\Gamma_1=\{m\in\Cset^2,\|m\|\leq {m_0\over 2}\}$, $\Gamma_2=\{m\in\Cset^2,{m_0\over 2}<\|m\|<m_0\}$ and $\Gamma_3=\{m\in\Cset^2,\|m\|\geq m_0\}$.

We end this section by a few remarks. First, one can do exactly the same computation with $\D_m=\D^0+\Delta(m)+J\Delta(m)J^{-1}$ where $\D^0$ is a fixed Dirac operator. Here one can see $\Delta(m)$ as a perturbation of the \og metric\fg\ $\D^0$ by the $1$-form $\Delta(m)$. The action is again a function of $m$ but with a translation from $m$ to $m-\mu$ where $\mu=m(\D^0)$ is the parameter associated to $\D^0$.

Gauge equivalent Dirac operators are related by $A\rightarrow A^u=uAu^*+u[D^0,u^*]$, where $u$ is a unitary element of $\pi(\A)$. The unitaries of $\pi(\A)$ are of the form $\pmatrix{U & 0\cr 0 & e^{i\theta}}$ where $U\in U(2)$. One can readily check that in terms of the parameter $m$ the gauge transformation $A\rightarrow A^u$ amounts to $m\rightarrow m^u=e^{-i\theta}U(m+\mu)-\mu$, where $U\in U(2)$. Thus the gauge equivalence  classes are the 3-spheres of $\Cset^2$ centered on $-\mu$. One can verify in this way that the spectral action is manifestly gauge invariant  (remember there is a translation by $-\mu$ if $\D^0\not=0$). The gauge group is easily seen to be $U(1)\times U(2)/U(1)=U(2)$.
 
\section{Quantization}

From the last section we see that the algebra of functions on phase space will be a direct product. Because of this we will perform canonical quantization on each connected component of the phase space and take the orthogonal direct sum of the three Hilbert spaces involved.

We write $\Gamma_a$ with $a=1,2,3$ and $m_{i,a},\bar m_{i,a}$ for the coordinates on $\Gamma_a$, with $i=1,2$. We have the symplectic structure :

\be
\Omega_a=i(dm_{1,a}\wedge\bar{dm_{1,a}}+dm_{2,a}\wedge\bar{dm_{2,a}})
\ee

and the canonical Poisson bracket relations :

\be
\{m_{i,a},m_{j,a}\}=\{\bar m_{i,a},\bar m_{j,a}\}=0,\qquad \{m_{i,a},\bar m_{j,a}\}=i\delta_{ij}
\ee

The canonical quantization is given by $m_{i,a}\rightarrow \hat{m}_{i,a}$, $\bar{m}_{i,a}\rightarrow \hat{\bar{m}}_{i,a}=\hat{m}_{i,a}^*$, and $\{\ ;\  \}\rightarrow -i[\ ,\ ]$ at least for the canonical Poisson bracket relations. The hatted $m_{i,a}$'s are operators on a Hilbert space. Thus one has the canonical commutation relations :

\be
[\hat m_{i,a},\hat m_{j,a}]=[\hat m_{i,a}^*,\hat m_{j,a}^*]=0,\qquad [\hat m_{i,a}, \hat m_{j,a}^*]=-\delta_{ij}
\ee

This is the usual algebra of two uncoupled quantum harmonic oscillators. In this case it is well known that there is a unique choice of Hilbert space representation up to unitary equivalence. We choose to represent $\hat m_{1,2}$ by creation operators and $\hat m_{1,2}^*$ by destruction operators. Thus we define the Hilbert space $\K_a$ with orthonormal basis $\{|s_a,t_a\ket,s,t\in\Nset\}$.

The annihilation and creation operators act by :

\bea
\hat m_{1,a}^*|s_a,t_a\ket=\sqrt{s}|s_a-1,t_a\ket\cr
\hat m_{2,a}^*|s_a,t_a\ket=\sqrt{t}|s_a,t_a-1\ket\cr
\hat m_{1,a}|s_a,t_a\ket=\sqrt{s+1}|s_a+1,t_a\ket\cr
\hat m_{2,a}|s_a,t_a\ket=\sqrt{t+1}|s_a,t_a+1\ket
\eea

with the convention that $|s_a,t_a\ket=0$ if $s$ or $t$ is negative. We also denote the vacuum state $|0_a,0_a\ket=|0_a\ket$. The basis ket $|s_a,t_a\ket$ is an eigenvector of the number operator defined by \goodbreak $N_a=\hat m_{1,a}\hat m_{1,a}^*+\hat m_{2,a}\hat m_{2,a}^*$, and we have :
\be
N_a|s_a,t_a\ket=(s+t)|s_a,t_a\ket
\ee

Now we define $\K=\K_1\buildrel{\perp}\over{\oplus}\K_2\buildrel{\perp}\over{\oplus}\K_3$. The operators $\hat{m}_{i,a}$ and $\hat{m}_{i,a}^*$ act trivially on $\K_b$ with $b\not=a$. Thus we have the canonical commutation relations :

\be
[\hat m_{i,a},\hat m_{j,b}]=[\hat m_{i,a}^*,\hat m_{j,b}^*]=0,\qquad [\hat m_{i,a}, \hat m_{j,b}^*]=-\delta_{ij}\delta_{ab}P_a\label{ccr}
\ee
where $P_a$ is the orthogonal projector on $\K_a$. The total number operator is $N=N_1+N_2+N_3$.

Now we must see how gauge transformations are implemented in $\K$. The gauge transformation $m\rightarrow m^u$ acts on $\K$, and this action is naturally decomposed by the eigenspaces of $N_a$. Indeed, the eigenspace of $N_a$ for the eigenvalue $n\in \Nset$ is spanned by $|s_a,t_a\ket={1\over \sqrt{s!t!}}\hat m_1^s\hat m_2^t|0_a\ket$ with $s+t=n$ and can be identified with the space of homogeneous polynomials of degree $n$ in the two variables $m_1,m_2$. On this space $u=(U,e^{i\theta})$ acts by the change of variable ${}^t m\rightarrow {}^t mU$ and an overall multiplication by $e^{in\theta}$. In fact if $U=e^{i\phi}U'$ with $U'$ in $SU(2)$, the $e^{i\phi}$ part will also contribute by a phase, so we can just look for the action of $U'\in SU(2)$. But we know that the space of homogeneous polynomials of degree $n$ in two variables is an irreducible unitary representation of $SU(2)$ of spin ${n\over 2}$. The unitarity is easy to see in the Bargmann representation where we identify vectors in $\K_a$ with $\Cset$-analytic functions on $\Cset^2$. The identification is given by :

\be
\sum_{s,t}a_{s,t}|s,t\ket=\sum_{s,t}{a_{s,t}\over \sqrt{s!}\sqrt{t!}}\hat m_1^s\hat m_2^t|0\ket\longmapsto( m\mapsto \sum_{s,t}{a_{s,t}\over \sqrt{s!}\sqrt{t!}} m_1^s m_2^t)\label{ident}
\ee

Note that we have dropped the index $a$ for simplicity. The first two sums are Hilbert sums, and we have :
\be
\sum_{s,t}|a_{s,t}|^2<\infty\label{l2}
\ee

The last sum in (\ref{ident}) is the power series representation of an analytic function which is defined on $\Cset^2$ thanks to (\ref{l2}). The scalar product in the Bargmann representation is given by :

\be
(f,g)={1\over \pi^2}\int_{\Cset^2}dm\bar f(m)g(m)e^{-\|m\|^2}\label{bargscal}
\ee 

We immediately see the $SU(2)$ invariance. Note that, because of the unicity of the extension of analytic functions, we are allowed to interpret the Bargmann representation of $\K_a$ as the space of restrictions to $\Gamma_a$ of entire functions on $\Cset^2$ such that (\ref{bargscal}) is finite.

Thus the scalar product is gauge-invariant, that is, $\K$ is a unitary representation of the gauge group.

\section{Quantization of the distance}
We first recall some definitions. In the following, $\A$ stands for any unital $C^*$-algebra, and $(\A,\H,\D)$ could be any spectral triple.
\begin{defn}
A state $s$ on $\A$ is any linear form such that
\begin{enumerate}
\item $s(a^*a)\geq 0$ (positivity)
\item $s(1_\A)=1$ (normalization)
\end{enumerate}
The space of all states on $\A$ is denoted by $\SS(\A)$. For all $s,s'\in\SS(\A)$, let $d(s,s')$ be :
\be
d(s,s')=\sup_{a\in\A}\{|s(a)-s(a')|, \|[\D,\pi(a)]\|\leq 1\}\label{connesdistform}
\ee
This defines a distance on $\SS(\A)$.\end{defn}

A state is called pure if it is not a convex combination of two other states. Several notions of spaces can be associated to a $C^*$-algebra, which all coincide when the algebra is commutative : the primitive spectrum, the structure space, and the space of pure states. The latter is appropriate for dealing with the distance defined above. For our case $\A=M_2(\Cset)\oplus\Cset$ the pure states are given by $p'(M,\alpha)=\alpha$ and $p_\xi(M,\alpha)=\xi^*M\xi$  where $\xi\in\Cset^2$ is normalized. Two normalized vectors give the same state {\it iff} they are colinear, that is {\it iff} their images in the complex projective line are the same, so the space of pure states of $\A$ is of the form $\SS_1\sqcup \SS_2$ where $\SS_1$ can be identified with  $\Cset P^1$ and $\SS_2=\{p'\}$ is a single point.

In \cite{kraj2} the authors compute the distance between the two pure states $p_\xi$ and $p'$. The result is :
\be
d(p_{\xi},p')=\|m\|^{-1}=(m_1\bar m_1+m_2\bar m_2)^{-1/2}\label{dist}
\ee

provided that $\xi$ and $m$ are colinear, which means that $p_\xi=p_{[m]}$ where we write $[m]$ for the class of $m$ in $\Cset P^1$. The distance is infinite if $\xi$ and $m$ are not colinear. In the appendix we give a different proof for this result and generalize it to a wider class of Dirac operators. See theorem \ref{ncdist}.

Let us emphasize that, exactly like in general relativity, the points, that is the pure states, have no physical meaning by themselves (except for one as we shall see). Now let us use \og general covariance\fg\  to rewrite (\ref{dist}) entirely in terms of the metric. Let us call $d_m$ the distance function determined by $\D_m$, and let $s,s'$ be any two points. Then one has :

\be
d_{m^u}(s,s')=d_m(s^u,s'^u)\label{cov}
\ee

where $u$ is any unitary of $\A$, playing here the role of an active diffeomorphism. It acts on a state $s$ by $s^u(a)=s(uau^{-1})$ so that $p_\xi^u=p_{u^{-1}\xi}$. Note that $p'$ is invariant under $u$. So one has :

\be
d(m):=d_m(p_{[m]},p')=d_{m^u}(p_{[m^u]},p')=\|m\|^{-1}\label{dist2}
\ee

This $d$, viewed as a function of the metric, is gauge-invariant. Thus it is a function of the geometry only (an observable), and it is this function that we want to quantize. Note that although a direct reference to the fixed point $p_{\xi}$ in the noncommutative \og manifold\fg\ has disappeared, as expected, the point $p'$ remains, since it has the uncanny property of being \og diff-invariant\fg. This is not surprising since it represents by itself a component of the manifold. So this particular point has a physical meaning. Note also that in this very simple model the geometry (that is entirely coded in $\|m\|$) is completely determined by the value of $d$. It entails that any classical observable will be a function of $d$.

Other types of distances can be computed, but they are always proportional to $d$, see \cite{kraj2} for details. In particular, if one identifies $\Cset P^1$ with the sphere $S^2$, such that $[m]$ is the north pole, then the distance between two antipodal points of the equator is $2d$. Thus one can naturally assign a radius $d$ to $\SS_1$. The whole situation can then be pictorially represented in $\Rset^3$ by the union of a sphere of radius $d$ and a point sitting at a distance $d$ above the north pole\footnote{However, one should not take this representation to the letter since the distance between two points of different lattitudes is infinite, as is the distance between the isolated point and any point of the sphere except the north pole. Nevertheless, the distance between two points lying at the same lattitude is the euclidean distance in $\Rset^3$. See \cite{kraj2} for details.}. So we can see $d$ as representing at the same time the distance between $\SS_1$ and $\SS_2$ and the radius of $\SS_1$.

Now we pass to the quantization $\hat d$ of $d$. The equations of motion, although trivial, enter the game here by selecting a different expression for $d$ in terms of the variables $m_{i,a}$, $\bar m_{i,a}$, according to which component of the phase space $m$ belongs to. On the component $\Gamma_a$ the expression for $d$ is $m_{1,a}\bar m_{1,a}+m_{2,a}\bar m_{2,a}$. The first term can be quantized as :
\be
(1-\alpha_a) \hat m_{1,a}\hat m_{1,a}^*+\alpha_a\hat m_{1,a}^*\hat m_{1,a}
\ee
for any real parameter $\alpha_a$ which captures the ordering ambiguities (e.g. $\alpha=1/2$ corresponds to the symmetric ordering). Thanks to the canonical commutation relations (\ref{ccr}) this reduces to $\hat m_{1,a}\hat m_{1,a}^*+\alpha_a P_a$. Similarly the second term is quantized as $\hat m_{2,a}\hat m_{2,a}^*+\beta_a P_a$. Thus the distance operator is given by :

\be
\hat d=(\sum_{i,a} \hat m_{i,a}\hat m_{i,a}^*+k_aP_a)^{-1/2}=(N+k_1P_1+k_2P_2+k_3P_3)^{-1/2}
\ee

where we introduce the constants $k_a=\alpha_a+\beta_a$. The result is meaningful for $k_1,k_2,k_3>0$. If we make the very natural assumption that the ordering problem is resolved in the same way on each component then $\hat d=(N+kI)^{-1/2}$, with $k=1$ for the symmetric ordering.  We see that distance is quantized, the eigenvalues form a discrete set and each one has a degeneracy (which might be viewed as unphysical, as we will see). More precisely, the spectrum is the set $\{(n+k)^{-1/2}|n\in\Nset\}$ in Planck units, and the eigenvalue $(n+k)^{-1/2}$ has multiplicity $3(n+1)$. In particular the underlying quantum noncommutative space has a maximum radius. In \cite{rovelli} Rovelli found the spectrum $(2n+k)^{-1/2}$, with $k=1$ for the symmetric ordering, and no multiplicity. The factor of $2$ in front of $n$ in Rovelli's formula comes from his equations of motion which imply a relation between $m_1$ and $m_2$ that we do not have. On the other hand, we get a non-trivial degeneracy structure which has two origins. The factor of $3$ in the multiplicity comes from the structure of the phase space (thus from the equations of motion), and the factor $n+1$ from the action of the gauge group. We will come back to this matter in the next section. 

We see $\hat d$ commutes with the unitary representation of the gauge group, and so it is a gauge-invariant observable, as expected from the classical analysis. As we said above, any other gauge-invariant classical observable is a function of $d$, and we expect that its quantization will be a function of $\hat d$. This is in fact easy to see since any self-adjoint operator $O$ commuting with the unitary representation of the gauge group will be a scalar multiple of the identity on every eigenspace of $N_a$ by Schur's lemma\footnote{The quantization of a function on the disconnected phase space $\Gamma$ will act separately on each subspace $\K_a$. A more general self-adjoint operator commuting with the action of the gauge group would be a direct sum of block hermitian $3\times 3$ matrices.}. That means that $O$ is of the form 

\be
O=\sum_{0\leq n<\infty\atop a=1,2,3} f_a(n)p_{a,n}\label{obs}
\ee
with $p_{a,n}$ the orthogonal projector on the eigenspace of $N_a$ for the eigenvalue $n$, and $f_a(n)\in\Rset$. This can be written $O=f(\hat d)$ with $f=(f_1,f_2,f_3)$ is a real function defined on the 3 components of the phase space which only depends on the radius. Thus we see that not only distance is quantized, but any other geometric quantity.

\section{Physical Hilbert space}

The Hilbert space $\K$ constructed above has two features which may be viewed as unphysical (of course the word \og unphysical\fg\ is to be taken with a grain of salt since this whole paper only deals with a mathematical toy model). In this section, which is admitedly speculative, we will try to get rid of these unphysical features.

First, $\K$ is decomposed as a direct sum $\K=\K_1\oplus\K_2\oplus\K_3$ where each summand corresponds to a component of the classical phase space. However, $\K_a$ contains eigenvectors of $N$ with eigenvalue outside the classically allowed region for $\|m\|^2$. Let us write $\K_a=\bigoplus_{n=0}^\infty \K^n_a$, the decomposition into the eigenspaces of $N$. Then we propose that the \og physically allowed part\fg\  of $\K$ is :

\be
\K'=\bigoplus_{0\leq n<{m_0^2/ 4}}\K_1^n\oplus\bigoplus_{{m_0^2/4}<n<m_0^2}\K_2^n\oplus\bigoplus_{n>m_0^2}\K_3^n\label{defofsubspace}
\ee

The space $\K'$ is a closed subspace of $\K$ which is invariant under $N$, with the same spectrum, but a degeneracy divided by $3$. Thus, all observables are well defined on $\K'$, and only their degeneracy changes. But of course there is an ambiguity in this procedure since $N+k_1P_1+k_2P_2+k_3P_3$ is an equally possible quantization of $\|m\|^2$ for any values of the $k_i$'s. If we take $k_1=k_2=k_3=k$, then $n$ becomes $n+k$ in (\ref{defofsubspace}) and this only amounts to a shift of the subspaces in the sum as long as $k$ is positive. Observables would have the same spectrum and degeneracy on the resulting subspace. However, if we do not take $k_1=k_2=k_3$ the procedure ends up with a space on which the obervables have the same spectrum but not the same degeneracy. We must of course forbid that and choose $0<k_1=k_2=k_3=k$. Fortunately this is precisely the choice we already made in the definition of $\hat d$. Another procedure, which do not suffer from this ordering ambiguity, is suggested by the Bargmann representation. In this representation, a general vector of $\K$ is of the form $(f_1,f_2,f_3)$ where the $f_i's$ are entire functions, as we already noted. If we put a continuity condition on the boundary of $\Gamma$ it prescribes $f_1=f_2=f_3$ (by the maximum principle for instance). This defines a space $\K''$ which is isomorphic to $\K'$, invariant by $N$, and on which $N$ has the same spectrum and the same degeneracy. In the following we assume that we have used one of these procedures, for instance the last one, to select a subspace that we will denote $\K$ again, and call the kinematical Hilbert space.

The second annoying fact that we want to get rid of is that the gauge tranformations only act as symmetries of the Hilbert space $\K$. There should exist a physical Hilbert space $\H_{phys}$ whose states (i.e. lines) entirely describe the physical system without mathematical redundancy. We look for a description of $\H_{phys}$ as well as a map $\Psi : \K\longrightarrow \H_{phys}$. We call $\K^n$ the eigensubspace of $\K$ corresponding to the eigenvalue $n$ of $N$, and if $v\in\K$ we write $v=\sum_{n=0}^\infty v^n$ the corresponding decomposition.

We propose that the map $\Psi$ satisfies the following conditions, for all $v\in\K$ :

\begin{enumerate}
\item $\Psi(\Cset v)\subset \Cset\Psi(v)$, which means that a kinematical state corresponds to a single physical state, 
\item $\forall u\in SU(2)$, $\Psi(u.v)\in\Cset \Psi(v)$, since gauge related kinematical states represent the same physical state,\label{cond2}
\item $\forall u\in SU(2)$, $\Psi(u.v+v)\in \Cset\Psi(v)$, since $u.v+v$ is the superposition of physically equivalent states. \label{cond3}
\end{enumerate}

We can already see that condition \ref{cond2}, together with the fact that $SU(2)$ acts irreducibly on $\K^n$, forbid $\Psi$ to be linear. Indeed, a linear $\Psi$ would require that all the vectors in the orbit ${\mathcal O}(v)$ under $SU(2)$ be linearly independent in order to be consistent with condition \ref{cond2}. It is also easy to see that conditions \ref{cond2} and \ref{cond3} are equivalent to :
$$\Psi(\mbox{Span}({\mathcal O}(v))\subset\Cset\Psi(v)$$

There are two more conditions that $\Psi$ (and the scalar product on $\H_{phys}$) must satisfy :

\begin{enumerate}
\setcounter{enumi}{3}
\item If $v^i\in\K^i$ and $v^j\in\K^j$ with $i\not=j$, then $\Psi(v_i)\perp\Psi(v_j)$,
\item\label{cond5} Let $v^i$ and $v^j$ be as above, and $a,b\in\Cset$. Suppose also that $\|v^i\|=\|v^j\|=1$ and $|a|^2+|b|^2=1$. Then $s=\Psi(av^i+bv^j)$ must satisfy :
$${|\bra s|\Psi(v^i)\ket|\over \|s\|\|\Psi(v^i)\|}=|a|,\ {|\bra s|\Psi(v^j)\ket|\over \|s\|\|\Psi(v^j)\|}=|b|$$
\end{enumerate}

This conditions are necessary in order to follow the general principles of quantum mechanics : two states in which the observable $N$ take different values must be orthogonal, and they can be superposed. The probabilistic interpretation must be the same at the kinematical and the physical level, and this is precisely what condition \ref{cond5} means. In fact we can even generalize this condition to a superposition of an arbitrary number of eigenstates, and we will need to do so below.

By the above conditions, $\H_{phys}$ must contain orthonormal vectors $(\epsilon^i)_{i\in\Nset}$, which are eigenvectors of $N$. We ask as a last minimality condition that this orthonormal system is a Hilbert basis for $\H_{phys}$. Then by condition \ref{cond5}, we must have for all $v^i\in\K^i, v^j\in\K^j$, $\Psi(v^i+v^j)=f^i(v^i)\epsilon^i+f^j(v^j)\epsilon^j$, with $|f^i(v^i)|=\|v^i\|$. Using the generalization of this condition to arbitrary sums of this type  we get for all $v\in\K$ :
\be
\Psi(v)=\Psi(\sum_{i=0}^\infty v^i)=\sum_{i=0}^\infty f^i(v^i)\epsilon^i
\ee
where the $f^i$'s are some functions which must satisfy $|f^i(v^i)|=\|v^i\|$. Thus $\Psi$ is uniquely determined up to arbitrary phases. We can take for instance :
\be
\Psi(v)=\Psi(\sum_{i=0}^\infty v^i)=\sum_{i=0}^\infty\|v^i\|\epsilon^i
\ee
The physical Hilbert space is also determined, but abstractly so : it is the Hilbert space spanned by the eigenvectors $\epsilon^i$. All the observables are well defined on it, with the same spectrum as on $\K$, but without degeneracy. At this point, we might want a more concrete representation of $\H_{phys}$. We can look for gauge-invariant states in the Bargmann representation. These would be $SU(2)$-invariant functions on $\Cset^2$, that is, functions of the form $f(\|m\|^2)$. Of course, except for the constant ones these cannot be $\Cset$-analytic and we must extend our space to find solutions. We also want our function space to be a Hilbert space generated by eigenvectors of $N$. In the Bargmann representation we have $N=m_1{\partial \over \partial m_1}+m_2{\partial \over \partial m_2}$. Thus we set $g(m)=f(\|m\|^2)$ and look for the solutions of the differential equation :
$$N(g)=ng\Longleftrightarrow \|m\|^2f'(\|m\|^2)=nf(\|m\|^2),\ \forall m\in\Cset^2$$
The solution is $g(m)=f(\|m\|^2)=k \|m\|^{2n}$ where $k$ is a constant. So we will represent $\epsilon^i$ by the function $e^i : m\mapsto k_i\|m\|^{2i}$ where $k_i$ is a normalization factor. If we choose $k_i={1\over \sqrt{i!}}$ then the map $\sum a_i\epsilon^i\longmapsto \sum a_ie^i$, which takes a Hilbert sum to a power series, is well defined. Then $\H_{phys}$ is identified with the set of functions of the form $g(m)=\sum_{i=0}^\infty b_n\|m\|^{2n}$, where $\sum_{i=0}^\infty n!|b_n|^2<\infty$. However, this identification is not canonical since we had to make a choice of normalization factors. Unfortunately we are not aware of a more natural interpretation of $\H_{phys}$. 

\section{Outlook and puzzles}

The example we have studied is well-behaved because there exists a simple, if not polynomial, relation between $\|m\|$ and $d$ and the canonical quantization procedure naturally equips us with a number operator which can be viewed as a quantization of $\|m\|^2$, up to a constant. Can we hope that everything will go that well when we come to models of physical interest ? In fact, already in the first natural generalization of our model, which is to consider the configuration space of all Dirac operators (\ref{genedirac}), the formula for $d$ in terms of $\|m\|$ and $\|n\|$ promises to be quite involved, since in the case where $[m]=[n]$ it already contains a supremum (see theorem \ref{ncdist}). We can also consider the issue of the eigenvalues of the Dirac operator, since every observable is computable in terms of them. In our nice example, the parameter $\|m\|$ is, up to a sign, the only non-trivial eigenvalue of the Dirac operator and this can also be seen as the reason for our success. What would happen for less nice examples ? For simplicity of exposition, let us collectively call $\lambda$ the eigenvalues of the Dirac operator and $m$ the parameter directly appearing in it. As the spectral triple grows in size, the computations of $\lambda$ in terms of $m$ would certainly turn out to be untractable. Thus we may be leaded to think that the computation of observables like $d$ out of $m$ would also be untractable, wheras it would be simpler in terms of $\lambda$. Thus, the $\lambda$ would possibly provide a better choice of coordinate on phase space than $m$ (see \cite{rl}). They would also ensure gauge invariance before quantization. This is not necessarily a good thing. For instance in the case at hand we would end up with just one real parameter $\|m\|$ and there would not be any symplectic structure. Moreover, in the general case the eigenvalues $\lambda$ have complicated relations between them. The question of which method is best in general will have to wait for future investigations.

However, there is another approach, also pioneered in \cite{rovelli}. Since the classical formulae for the distance, or maybe other geometric quantities, may lead to analytical difficulties, we might be satisfied with a direct quantization of the Dirac operator. In \cite{rovelli} the author defines $\hat\D$ on $\hat\H:=\H\otimes\K$ in the following way. First we identify $\hat\H$ with $3\times 3$ matrices with entries in $\K$, then we define $\hat A\in\H\otimes\B(\K)$ by 
\be
\hat A=\pmatrix{0&0&\hat m_1\cr 0&0&\hat m_2\cr \hat m_1^*&\hat m_2^*&0}
\ee
and for any decomposed tensor $\psi\otimes\phi\in\hat\H$ we set
\be
\hat\D(\psi\otimes\phi)=\hat A(\phi)\psi+\psi\hat A(\phi)\label{quantumdirac}
\ee
where 
\be
\hat A(\phi)=\pmatrix{0&0&\hat m_1(\phi)\cr 0&0&\hat m_2(\phi)\cr \hat m_1^*(\phi)&\hat m_2^*(\phi)&0}\in\hat\H
\ee
and the product in (\ref{quantumdirac}) is the matrix product.
For any normalized $\phi\in\K$, which represents a quantum geometry, we have a projection :
\bea
\bra \phi|.\ket : \hat\H&\longrightarrow&\H\cr
\psi\otimes\xi&\longmapsto&\psi\bra\phi|\xi\ket
\eea
and a Dirac operator $\D_\phi$ on $\H$ defined by\footnote{We depart a little from Rovelli's analysis which he has carried entirely within $\H_\phi=\H\otimes\phi\simeq\H$, but it amounts to the same for our purpose.} 
\be
\D_\phi(\psi)=\bra\phi|\hat\D(\psi\otimes\phi)\ket,\quad \forall\psi\in\H
\ee
From this point of view, the expectation value $d^\phi$ of $d$ in the state $\phi$ should be given by Connes' distance formula (\ref{connesdistform}) with the Dirac operator $\D_\phi$. The main advantage of $d^\phi$ compared with $\bra \phi|\hat{d}|\phi\ket$ is that we do not need to have solved $d$ in terms of $m$ to get a formula in the quantum setting. Another is that we do not have ordering ambiguities. But do the two values agree ? It is easy to see that in our example :
\be
d^\phi=(\bra\phi|\hat{m_1}|\phi\ket\bra\phi|\hat{m_1}^*|\phi\ket+\bra\phi|\hat{m_2}|\phi\ket\bra\phi|\hat{m_2}^*|\phi\ket)^{-1/2}\label{dphi}
\ee
This value is clearly different from $\bra\phi|\hat{d}|\phi\ket$ : for instance in a state $|s,t\ket$, which is an eigenstate of $\hat d$, (\ref{dphi}) gives an infinite result. The difference clearly comes from having taken the expectation value first in one hand, and last in the other hand. Which way is correct we do not know. 

So it seems we have raised more questions than we have answered. It is only by studying more complex examples and comparing the different approaches exposed in the end that we will get some insight. For the moment we will be happy if we have succesfully exposed the main issues and motivated more work along that lines.

\ack

We wish to thank Mark Hale for comments on a preliminary version of this paper, and the anonymous referee for suggesting some clarifications.
\appendix
\section{Dirac operators on $(\A,\H,J,\chi)$}

The fixed structures are the algebra $\A=M_2(\Cset)\oplus\Cset$, the Hilbert space $\H=M_3(\Cset)$ together with the representation $\pi$ of $\A$ :
\bea
\pi(A\oplus a)&:&\H\longrightarrow \H\cr
& & \psi\longmapsto  \pmatrix{A&0\cr 0&a}\psi\label{repr}
\eea
the charge conjugation :
\bea
J&:&\H\longrightarrow \H\cr
& &\psi\longmapsto \psi^*\label{conj}
\eea
and finally the chirality :
\bea
\chi&:&\H\longrightarrow \H\cr
& &\psi\longmapsto \psi_\gamma\psi\psi_\gamma\label{par}
\eea
where $\psi_\gamma=\pi(I_2\oplus -1)=\pmatrix{1&0\cr 0&-1}$. We are looking for all Dirac operators $\D$ such that $(\A,\H,\D)$ ($\chi$, $\pi$, $J$ are understood) satisfy the axioms of a finite spectral triple. It turns out that $\H$, $\pi$, $J$ and $\pi$ are entirely determined by a symmetric matrix $\mu\in M_2(\Zset)$ called the multiplicity matrix. In our case :

\be
\mu=\pmatrix{1&-1\cr -1&1}
\ee

Since it is degenerate, the Poincar\'e duality axiom will not be fulfilled, but we ignore this point. Let us explain on this example how we recover $\H$, $\pi$, $J$ and $\chi$ from $\mu$ (further details can be found in \cite{kraj}). We define $\H$ first :
\be
\H=\bigoplus_{1\leq i,j\leq 2}\H_{ij},\mbox{ with }\H_{ij}=\Cset^{n_i}\otimes\Cset^{|\mu_{ij}|}\otimes\Cset^{n_j}\label{decomp}
\ee
where $n_1=2$ and $n_2=1$. This corresponds to the dimension of the fundamental representation of $M_2(\Cset)$ and $\Cset$ respectively.

Since $|\mu_{ij}|=1$ we have :
\be
\H=\Cset^2\otimes\Cset^2\oplus\Cset^2\otimes\Cset\oplus\Cset\otimes\Cset^2\oplus\Cset\otimes\Cset
\ee
where the summand are mutually orthogonal and endowed with the usual hermitian product. Charge conjugation is defined by :
\bea
J &:& \H_{ij}\longrightarrow \H_{ji}\cr
& &x\otimes y\longmapsto \bar y\otimes\bar x
\eea

The representation $\pi$ is defined in the following way :
\bea
\pi(a_1\oplus a_2) &:& \H_{ij}\longrightarrow \H_{ij}\cr
 & & x\otimes y\longmapsto a_ix\otimes y
\eea

There is also a representation on the right defined by $J\pi J^{-1}$. Thus if we identify $\Cset^{n_i}\otimes\Cset^{n_j}$ with $M_{n_i,n_j}(\Cset)$ by $x\otimes y\mapsto {}^txy$, then $\A$ acts on the left by matrix multiplication on column vectors and on the right by matrix muliplication on row vectors. In this way, $\H$ is identified with $M_3(\Cset)$, and the decomposition $\H=\bigoplus_{ij}\H_{ij}$ corresponds to the block matrix decomposition :
\be
\H\ni\psi=\pmatrix{\psi_{11}&\psi_{12}\cr \psi_{21}&\psi_{22}}
\ee
with $\psi_{11}\in M_2(\Cset)$, $\psi_{12}\in M_{2,1}(\Cset)$, $\psi_{21}\in M_{1,2}(\Cset)$, $\psi_{22}\in\Cset$. One can then readily verify that $\pi$, $J$ and $\chi$ are given by (\ref{repr}), (\ref{conj}), (\ref{par}).

It is proven in \cite{kraj} that a Dirac operator on $\pi$, $J$, $\chi$ can be uniquely decomposed in the form :
\be
\D=\Delta+J\Delta J^{-1}
\ee
where $\Delta$ is linear on the right ($\Delta(\psi.a)=\Delta(\psi).a$ forall $\psi\in\H$ and $a\in\A$), hermitian and anticommutes with $\chi$. Furthermore, if we define the projection operators $P_{ij} : \H\rightarrow \H_{ij}$ given by the decomposition (\ref{decomp}), then the matrix elements $\Delta_{ij}^{kl}=P_{ij}\Delta P_{kl}$ are of the form :
\be
\Delta_{ij}^{kl}=\delta_{jl}M_{ik,j}\otimes I_{n_j}
\ee
with $M_{ik,j}\in M_{|\mu_{ij}|n_i,|\mu_{kj}|n_k}(\Cset)$. If $\mu_{ij}\mu_{kj}<0$ then $M_{ik,j}=M_{ki,j}^*$, else $M_{ik,j}=0$. In the case at hand, this constraints $\Delta$ to have to following form :
\be
\Delta=\matrix{
 \matrix{\H_{11} &\qquad\H_{12}  &\qquad \H_{21} &\qquad \H_{22}} & \cr
\pmatrix{0&0&M_{12,1}\otimes I_2&0\cr 0&0&0&M_{12,2}\cr M_{12,1}^*\otimes I_2&0&0&0\cr 0&M_{12,2}^*&0&0}&\matrix{\H_{11}\cr \H_{12}\cr \H_{21}\cr \H_{22}}
}
\ee
Since $M_{12,1}\in M_{2,1}(\Cset)$ we can define $m=M_{12,1}=\pmatrix{m_1\cr m_2}$. Similarly we note $n=M_{12,2}=\pmatrix{n_1\cr n_2}$. We also define $A(m)=\pmatrix{0&m\cr m^*&0}$ and similarly for $A(n)$. Then $\Delta$ has the following interpretation in terms of matrix multiplication :

\be
\Delta(\psi)=A(m)\psi E_1+A(n)\psi E_2
\ee
where $E_1=\pi(I_2\oplus 0)$ and $E_2=\pi(0\oplus 1)$. Thus, the complete and most general Dirac operator $\D$ such that $(\A,\H,\D)$ is a finite spectral triple is :
\be
\D(\psi)=A(m)\psi E_1+A(n)\psi E_2+E_1\psi A(m)+E_2\psi A(n)\label{genedirac}
\ee
If $m=n$ then $\D(\psi)=A(m)\psi+\psi A(m)$ and it this sector of the theory that we consider in this paper.

Other cases are also of interest. In fact, using the results in \cite{kraj} it is easy to see that if, and only if, $m$ and $n$ are colinear, then the triple $(\A,\H,\D)$ comes from a gauge theory. More precisely, this gauge theory is defined on $(\A_0,\H_0,\D_0)$, with $\A_0=\Cset\oplus\Cset$, $\H_0=M_2(\Cset)$ on which $\A_0$ acts by diagonal matrices. Moreover, if we write $e_1=\pi(1\oplus 0)$ and $e_2=\pi(0\oplus 1)$ then $\D_0$ is of the form :
\be
\D_{x,y}(\phi)=a(x)\phi e_1+a(y)\phi e_2+e_1\phi a(x)+e_2\phi a(y)\label{genedirac0}
\ee
where $\phi\in M_2(\Cset)$, $x,y\in\Cset$ and $a(z)=\pmatrix{0&z\cr \bar z&0}\in M_2(\Cset)$. On this triple, we consider the module ${\mathcal E}=\Cset^2\oplus\Cset$ and a Hermitian connection $\nabla$ which depends on a vector $\mu\in\Cset^2$. Then one recovers $(\A,\H,\D)$ by the formulae given in \cite{kraj} and one finds that $m=x\mu$, $n=y\mu$. Thus, the spectral triple $(\A,\H,\D)$ in the case $m=n$ that we consider in this paper can be viewed as a noncommutative vector bundle on the two-point noncommutative manifold $(\A_0,\H_0,\D_0)$ with
\be
\D_0(\phi)=a(x)\phi+\phi a(x)^*
\ee

\section{Computation of the distance}

In this appendix we wish to compute the distance $d(p_\xi,p')$ for the Dirac operator (\ref{genedirac}).
\begin{lem}
We have for all $a\in\A$ :
\be
\|[\D_{m,n},\pi(a)]\|_{\B(\H)}=\sup(\trip [A(m),\pi(a)]\trip,\trip [A(n),\pi(a)]\trip)
\ee
\end{lem}
\pf
First let us explain the notations. For any $T\in\B(\H)$, $\|T\|_{\B(\H)}=\sup_{\|\psi\|=1}\|T\psi\|$, where the norm of $T\psi$ is the hermitian norm in $\H$. On the other hand, for any $M\in M_3(\Cset)$ we write $\trip M\trip=\sup_{\|X\|=1}\|MX\|$, where $X\in\Cset^3$ and the norm is the usual hermitian norm on $\Cset^3$.
Now we have, for all $a\in\A$ :
\be
[\D_{m,n},\pi(a)]\psi=[A(m),\pi(a)]\psi E_1+[A(n),\pi(a)]\psi E_2
\ee
Let us write $\psi=\pmatrix{X_1&X_2&X_3}$ where $X_i\in\Cset^3$, and set $A=[A(m),\pi(a)]$, $A'=[A(n),\pi(a)]$. Then we have :
\be
[\D_{m,n},\pi(a)]\psi=\pmatrix{AX_1&AX_2&A'X_3}
\ee
It is clear then that $\|[\D_{m,n},\pi(a)]\|_{\B(\H)}\geq \sup(\trip A\trip,\trip A'\trip)$. To have the other inequality it suffices to choose a matrix $\psi$ of norm $1$ that realizes the supremum of $\|[\D_{m,n},\pi(a)]\psi\|$. Then we have :
\bea
\|[\D_{m,n},\pi(a)]\|_{\B(\H)}^2&=&\|AX_1\|^2+\|AX_2\|^2+\|A'X_3\|^2\cr
&\leq &\sup(\trip A\trip,\trip A'\trip)^2(\|X_1\|^2+\|X_2\|^2+\|X_3\|^2)
\eea
This ends the proof.\qed

\begin{lem}
We have :
\be
d(p_\xi,p')=\sup\{\xi^*z\xi\ |\  z\in M_2(\Cset), z^*=z,\|zm\|\leq 1,\|zn\|\leq 1\}
\ee

\end{lem}
\pf First it is known \cite{kraj2} that the supremum can be taken over self-adjoint $a\in\A$. In that case if $a=(A,\alpha)$ it is an easy calculation that for all $X\in\Cset^2$, $y\in\Cset$ :
\be
[A(m),\pi(a)]{\pmatrix{X\cr y}\atop }={\pmatrix{(\alpha-A)my\cr \bra (A-\alpha)m|X\ket}\atop }
\ee
Then one has :
\be
\sup_{\|X\|^2+|y|^2=1}\|(\alpha-A)m\|^2|y|^2+|\bra (A-\alpha)m|X\ket|^2=\|(A-\alpha)m\|^2
\ee
Finally we observe that $p_\xi(a)-p'(a)=\xi^*A\xi-\alpha\xi^*\xi=\xi^*z\xi$ where we set $z=A-\alpha$. \qed

We write $H_2$ for the set of $2\times 2$ hermitian matrices. We will use the basis given by Pauli matrices :
\be
\sigma^0=\pmatrix{1&0\cr 0&1},\ \sigma^1=\pmatrix{0&1\cr 1&0},\ \sigma^2=\pmatrix{0&-i\cr i&0},\ \sigma^3=\pmatrix{1&0\cr 0&-1}
\ee

What we have to do is to extremize the linear form $\phi_\xi(z)=\xi^*z\xi$ on the set $C=C_m\cap C_n$ where $C_m : \|zm\|^2\leq 1$, $C_n : \|zn\|^2\leq 1$. It turns out that $C_m$ and $C_n$ are two (solid) cylinders. Let $B_m$ (resp. $B_n$) be the bilinear form on $H_2$ defined by polarization of $\|zm\|^2$ (resp. $\|zn\|^2$). For all $h,h'\in H_2$ :
\bea
B_m(h,h')&=&{1\over 2}(\bra hm|h'm\ket+\bra h'm|hm\ket)\cr
&=&{1\over 2}(\bra m|(hh'+h'h)m\ket)\cr
&=&{1\over 2}\bra m|\{h,h'\}m\ket
\eea
where we have introduced the Jordan bracket in the last line. The Pauli matrices satisfy :
\be
\{\sigma^i,\sigma^j\}=2\delta_{ij}
\ee
for $i=1,2,3$. Since $B_m(\sigma^0,\sigma^i)=\bra m|\sigma^im\ket$, in the basis of Pauli matrices the matrix of $B_m$ (that we also call $B_m$) is :
\be
B_m=\pmatrix{\|m\|^2&\bra m|\sigma^1m\ket&\bra m|\sigma^2m\ket&\bra m|\sigma^3m\ket\cr \bra m|\sigma^1m\ket&\|m\|^2&0&0\cr \bra m|\sigma^2 m\ket&0&\|m\|^2&0\cr \bra m|\sigma^3m\ket&0&0&\|m\|^2}\label{bil}
\ee
$B_m$ has the form $\|m\|^2I_4+S$, where $S$ is obviously of rank $2$ and has zero trace. Thus there exists an orthonormal matrix $O$ and a non-negative real $\lambda$ such that :
\be
B_m=O^\top\pmatrix{\|m\|^2-\lambda&0&0&0\cr 0&\|m\|^2+\lambda&0&0\cr 0&0&\|m\|^2&0\cr 0&0&0&\|m\|^2}O
\ee
To find $\lambda$ we just need the determinant of $B_m$ which turns out to be zero by (\ref{bil}). Thus $\lambda=\|m\|^2$ and it follows that $C_m$ is a cylinder with ellipsoid basis. Now we are looking for the eigendirections of $C_m$.

\begin{defn}
For any $u\in H_2$ we write $u=\pmatrix{u_0\cr \vect{u}}$. We write $.$ for the scalar product in the hyperplane $W$ generated by $\sigma^i, i=1,2,3$. We also define the map $\vect{h} : \Cset^2\setminus\{0\}\longrightarrow W$ by 
\be
\vect{h}_m={\displaystyle{1\over\|m\|^2}}\pmatrix{\bra m|\sigma^1m\ket\cr \bra m|\sigma^2m\ket\cr\bra m|\sigma^3m\ket}
\ee
\end{defn}

The map $\vect{h}_m$ is actually none other than the Hopf fibration\footnote{Strictly speaking it is the Hopf fibration when retricted to any $3$-sphere.} as can be seen through the following calculation :
\bea
\bra m|\sigma^1m\ket&=&\bar m_1m_2+m_1\bar m_2\cr
\bra m|\sigma^2m\ket&=&-i(\bar m_1m_2-m_1\bar m_2)\cr
\bra m|\sigma^3m\ket&=&|m_1|^2-|m_2|^2
\eea
In the sequel we will write $[m]$ for the class of $m$ in $\Cset P^1$, and we continue to denote by $\vect{h}$ the identification between $\Cset P^1$ and $S^2\subset W$ given by the Hopf fibration.

Let $\vect{u}_m$ be a vector in $W$ orthogonal to $\vect{h}_m$ and $\vect{u}_m'=\vect{h}_m\times\vect{u}_m$. Then it is an easy exercise to verify that :
\bea
V_0(m)=\pmatrix{-1\cr \vect{h}_m},& V_2(m)=\pmatrix{1\cr\vect{h}_m}\cr
V_1(m)=\pmatrix{0\cr \vect{u}_m},& V_1'(m)=\pmatrix{0\cr \vect{u}_m'}\label{eigendir}
\eea
are eigenvectors for $B_m$, for the eigenvalues $0$, $2$, $1$, $1$ respectively. Thanks to the above we see immediately that $C_m$ and $C_n$ have parallel axis if and only if $[m]=[n]$. In the same vein we have :

\begin{lem}
The linear form $\phi_\xi$ is bounded on $C_m$ if and only if $[\xi]=[m]$. In that case $\sup_{C_m}\phi_\xi(z)=\|m\|^{-1}$.
\end{lem}
\pf First we observe that $\phi_\xi(\sum_\mu z_\mu\sigma^\mu)=\sum_\mu\bra \xi|\sigma^\mu\xi\ket z_\mu=\bra V_2(\xi)|z\ket$ where $V_2(\xi)$ is defined as in (\ref{eigendir}). Now $\phi_\xi$ is bounded on $C$ if and only if its kernel is parallel to the axis of the cylinder, that is, iff $V_2(\xi)\perp V_0(m)$. But by Cauchy-Schwarz this is equivalent to $\vect{h}_\xi=\vect{h}_m$, and finally to $[\xi]=[m]$. 
Now if $[\xi]=[m]$ then $V_2(\xi)=V_2(m)$ thus it is clear that the supremum of $\phi_\xi$ is reached at a summit $S(m)$ of the ellipsoid basis of $C_m$ in the direction $V_2(m)$. We write $S(m)=\lambda V_2(m)$. Using $B_m(S(m),S(m))=1$ we find $\lambda^2={\displaystyle{1\over 4\|m\|^2}}$. The supremum will in fact correspond to the positive square root, so we obtain finally :
\be
\sup_{C_m}\phi_\xi(z)=\phi_\xi(\lambda V_2(m))={1\over 2\|m\|}\|V_2(m)\|^2={1\over\|m\|}
\ee
This concludes the lemma. \qed

From this lemma we get 

\begin{thm}\label{ncdist}
If $[m]=[n]$ then the $d(p_\xi,p')$ is finite if and only if $[\xi]=[m]=[n]$ and its value is sup$(\|m\|^{-1},\|n\|^{-1})$.
\end{thm}
\pf This clearly follows from the lemma since $B_m=\alpha^2 B_n$ and $C_m=\alpha C_n$ with $\alpha=\|m\|/\|n\|$.\qed

So we have recovered in a different way and slightly generalized the formula given in \cite{kraj2}. The generalization to the case $[m]\not=[n]$ is not known to us yet.

\end{document}